\documentclass[final,nopageno]{jpaper}

\usepackage[normalem]{ulem}
\usepackage{threeparttable}
\usepackage{amsmath,bm}

\begin{document}

\title{
Twin-Load: Building a Scalable Memory System over the Non-Scalable Interface
}

\author{
  Zehan Cui$^{1,2}$\quad Tianyue Lu$^{1,2}$\quad Haiyang Pan$^{1,2}$\quad Sally A. Mckee$^{3}$\quad Mingyu Chen$^{1*}$\\
  ${}^{1}$Insitute of Computing Technology, Chinese Academy of Sciences,Beijing,Chin\\
  ${}^2$University of Chinese Academy of Sciences,  Beijing,China\\
  ${}^3$Chalmers University of Technology,Sweden\\
  \authemail{\{cuizehan,lutianyue,panhaiyang\}@ict.ac.cn mckee@chalmers.se cmy@ict.ac.cn}
}

\date{}
\maketitle

\thispagestyle{empty}

\begin{abstract}

Commodity memory interfaces have difficulty in scaling memory capacity
to meet the needs of modern multicore and big data systems.
DRAM device density and maximum device count are constrained by technology, package, 
and signal integrity issues that limit total memory capacity.
Synchronous DRAM protocols require data to be returned within a fixed latency,
and thus memory extension methods over commodity DDRx interfaces 
fail to support scalable topologies. Current extension approaches
either use slow PCIe interfaces, 
or require expensive changes to the memory interface, which limits
commercial adoptability.

Here we propose \textbf{twin-load}, a lightweight
asynchronous memory access mechanism over the synchronous DDRx interface.
Twin-load uses two special loads to accomplish one access request to 
extended memory --- the
first serves as a prefetch command to the DRAM system, 
and the second asynchronously gets the required data. 
Twin-load requires no hardware changes on the processor side 
and only slight software modifications.
We emulate this system on a prototype to demonstrate the feasibility of
our approach. 
Twin-load has comparable performance to NUMA extended memory
and outperforms a page-swapping PCIe-based system by 
several orders of magnitude.
Twin-load thus enables instant capacity increases on commodity platforms,
but more importantly, our architecture opens opportunities
for the design of novel, efficient, scalable, cost-effective memory subsystems.

\end{abstract}

\section{Introduction}

Commodity memory interfaces have difficulty in scaling memory capacity
to meet the needs of modern multicore and big data systems. For instance,
the number of cores in chip multiprocessors
(CMPs) is growing such that memory capacity per core drops by 30\%
every two years~\cite{lim2009disaggregated}. In virtualized environments,
running many consolidated virtual machines per core further increases
memory requirements~\cite{vmware16}.
Sufficient capacity to hold frequently used ``hot'' data 
becomes critical to avoid slow disk accesses. 
For example, the total memory used for data caching in Facebook is
about 75\% of the size of its non-image data~\cite{ousterhout2011case}. 
In-memory databases can perform queries 100 
times faster than traditional disk-base approaches~\cite{oracle2013,graefe2014memory}.
And Google, Yahoo, and Bing store their search indices entirely in 
DRAM~\cite{kozyrakis2010server}.

This ``capacity wall'' has several causes.
First, the number of channels is limited by the processor's pin count, 
estimated to increase only by 6.5\% each year~\cite{itrs2012}.
Second, the DRAM channel connects multiple dual in-line 
memory modules (DIMMs) via a multi-drop bus. Signal integrity (SI) 
requires that higher-frequency bus supports fewer DIMMs per channel, e.g., the newest Intel 
Xeon only supports one dual-rank DIMM per channel for DDR3-1866~\cite{rdimmcnt}.
Third, it is challenging to scale DRAM feature sizes below 20nm~\cite{itrs2012,nair2013archshield}.

%
%


Buffer chips can be used to mitigate pin count and SI 
limitations~\cite{cooper2012buffer,cisco,lrdimm}, but
the processor's maximum tolerable access latency still
restricts current solutions to one-layer extensions.
This constraint can be avoided by accessing memory via  
packet-based asynchronous 
protocols~\cite{hmc,udipi2011combining,lim2009disaggregated,lim2012system,hou2013cost}.
For instance, standard PCIe can be used to access DRAM in remote servers~\cite{hou2013cost}
and disaggregated memory blades~\cite{lim2009disaggregated,lim2012system}, but 
random accesses to large working sets can
suffer 100$\times$ slowdowns~\cite{hou2013cost}.
Asynchronous protocols can be implemented over high-speed serialized links~\cite{hmc} 
or photonics~\cite{udipi2011combining}, but the expense has 
thus far limited their 
widely adoption.
We believe that an industry-standard asynchronous interface is the right way to go,
But any solution that requires changes to the processor
interface slows commercial adoption.

%
%
We seek a practical solution to the capacity wall that
is much more scalable, requires no changes to commodity
processors and memory modules, and delivers acceptable
performance for big-memory applications. 
We use Memory Extending Chips (MECs) to build
a multi-layer memory system, the extra propagation delay for which violates 
DRAM latency constraints. To address this,
each access to the extended memory is replaced by two special \emph{twin-loads}, where
the first one prefetches data into the MEC buffer, and the second one 
brings it into the processor.
The twin-load addresses point to the same location, but
we manipulate them so that 1)~the second load reaches the MEC rather than hitting in cache
and 2)~the commodity processor is tricked into serializing them
on the memory interface to ensure time to complete the prefetch.

%
%
Our contributions are:
\begin{itemize}
\item
We implement an asynchronous protocol over the synchronous DDRx interface
by introducing a twin-load mechanism that coordinates software and hardware.
\item
We propose to use commodity processors and memory modules to create a scalable
extended memory system based on twin-load.
We study two different mechanisms 
to guarantee prefetch-to-load order and enough time for
the prefetch. One of them enables exploiting the memory concurrency.
\item
We implement a software prototype that reserves memory
from the operating system to emulate the extended memory and MECs, 
and we implement a lightweight extended 
memory manager. 
\item
We use our prototype to evaluate the feasibility of our proposals, finding that
our best solution has comparable performance with NUMA extensions, but better
scalability and performance per dollar.
For applications
with large working sets and irregular accesses, our twin-load solutions 
perform orders of magnitude better than page-swapping with PCIe-connected remote memory.
Even when compared to an ideal system with all local memory, twin-load
incurs an acceptable slowdown (about 26\%).
\end{itemize}

\noindent
The ability to quickly, easily, and inexpensively increase memory capacity
delivers good return-on-investment, but it is not the most compelling benefit
of our memory architecture. Rather, our approach opens new opportunities
to build innovative, efficient memory systems 
such as integrating remote memory pools, heterogeneous DRAM/NVM, direct remote memory access,
and even accelerators into the MECs. 

\section{Background and Related Work}

\label{section:protocol}


\begin{figure}[!t]
\centering
\includegraphics[width=7.50cm]{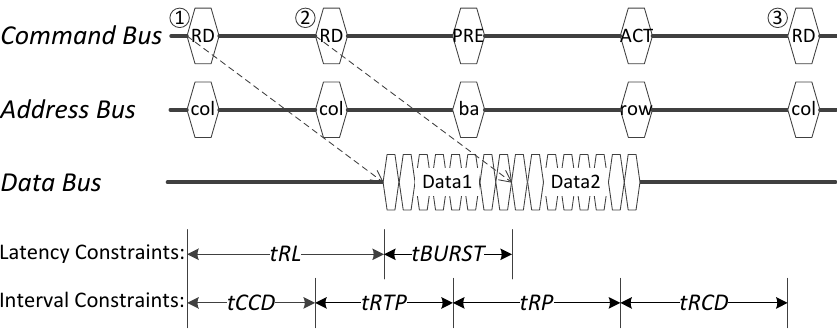}
\caption{DRAM Access Protocol. RD~\textcircled{1} and RD~\textcircled{2} map to the same bank and 
row, while RD~\textcircled{3} maps to another row in that bank. The rank and bank addresses 
associated with each command are omitted for readability. }
\label{figure:protocol}
\end{figure}

\begin{table}[!t]
\centering
{
  \caption{DDRx Timing Parameters}
\sffamily
\footnotesize
{
  \begin{threeparttable}
  \begin{tabular}{|c|l|c|}
    \hline
    Timing~\tnote{a} & Description & Typical \\
    Parameter & & Value \\
    \hline
    \hline
    $tRL$  	& Fixed latency from RD command to first data 	& 13.75$ns$ 	\\
    \hline
    $tBURST$ 	& Fixed duration of data transfer		& 4 cycles	\\
    \hline
    $tCCD$	& Minimum delay between two RD commands		& 4 cycles	\\
    \hline
    $tRTP$	& Minimum delay between RD and PRE commands	& 7.5$ns$		\\
    \hline
    $tRP$  	& Minimum delay between PRE and ACT commands 	& 13.75$ns$ 	\\
    \hline
    $tRCD$ 	& Minimum delay between ACT and RD commands 	& 13.75$ns$ 	\\
    \hline
  \end{tabular}
{
  \begin{tablenotes}
    \scriptsize
    \item[a] Delays between commands to the same bank 
  \end{tablenotes}
}
  \end{threeparttable}
  \label{table:timing}
}
}
\end{table}

DRAM memory systems usually consist of multiple channels
that drive DIMMs composed of multiple ranks,
and all ranks on a channel share a command, address, and data bus.
A rank contains multiple 
storage arrays, or banks.
Total memory capacity is thus determined by the number of channels,
the ranks per channel, and the rank capacity, which 
are limited by the pin count, signal integrity, and chip density, respectively.

Simple synchronous protocols like JEDEC DDRx~\cite{ddr3spec,ddr4spec} 
are commonly used to access the data arrays; in such protocols,
data are placed on the bus a fixed latency after being requested,
and thus no handshake occurs between the producer and consumer. 
An \emph{activate} (ACT) command ``opens'' a given row,
loading the target row's data into a bank of sense amplifiers from
which read (RD) and write (WR) commands access the data at specified columns.
Subsequent accesses to an open row (\emph{row hits}) require no
ACT command to resend the row address. 
When data from a different row are needed (\emph{row misses}), the memory controller sends
a \emph{precharge} (PRE) command to ``close'' the row, which
writes the data back to the storage array and precharges the bank's sense amplifiers.
It then issues an ACT command to open the new row.

Figure~\ref{figure:protocol} illustrates the basic operation, and
Table~\ref{table:timing} shows the related timing parameters.
For example, RD~\textcircled{2} in Figure~\ref{figure:protocol}
is a row hit. The ACT is omitted,
and the RD can be issued after a short $tCCD$ latency.
RD~\textcircled{3} is a row miss, and so
a PRE command is sent after $tRTP$ time to close the row.
After $tRP$ time to finish the precharge, an ACT command with the new 
row address is sent, and the RD command with the column address can finally 
be issued after $tRCD$ time.


\begin{figure}[!t]
\centering
\includegraphics[width=7.50cm]{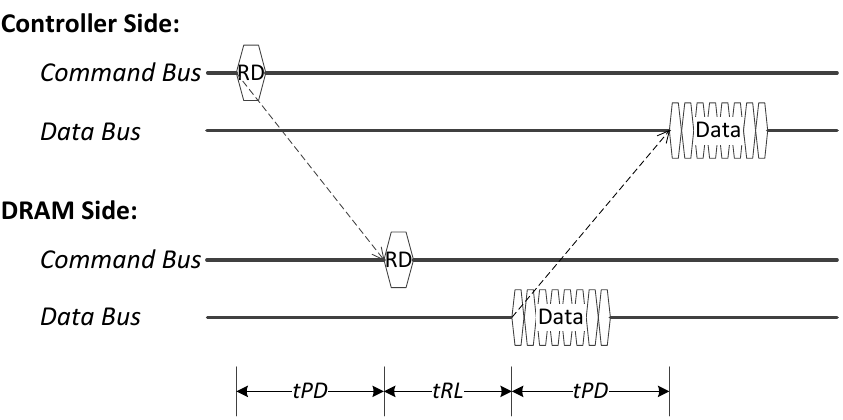}
\caption{DRAM Access with Extending Hardware. $tPD$ is the propagation delay
of commands/data between the memory controller and DRAM chips.}
\label{figure:latency}
\end{figure}


\subsection{Memory Extension over Standard DDRx Interfaces}

Buffer chips can be used to alleviate pin count and SI constraints within 
the latency requirements of DDRx protocols.
For instance, Cisco's unified computing system extended technology uses an 
ASIC buffer to expand a DDR3 DRAM channel into four 
distinct channels~\cite{cisco}, yielding up to 4$\times$ the capacity.
Such buffer chips with five DDRx interfaces become expensive due to
large die area and packaging costs.
LRDIMM uses a memory buffer to re-drive the DRAM bus, alleviating SI issues
and enabling more DIMMs/channel~\cite{lrdimm}.
Scalability is still limited at higher frequencies: e.g., at DDR3-1866
the newest Intel Xeon only supports one LRDIMM per channel~\cite{lrdimmcnt}.

Both methods have limited scalability because they can only support
one-layer extensions.
When extending hardware is interposed between the memory controller and DRAM chips,
commands and data experience extra propagation delays. 
For DRAM writes, commands and data propagate in the same direction, remaining
synchronous on the channel. DRAM reads require longer round-trip times because 
data and commands move in opposite directions.
Figure~\ref{figure:latency} shows the difference in read timing. 
For the simplest scenario where the extending hardware just forwards commands and 
data without any processing, the extra delay is 3.4$ns$ in each direction~\cite{lrdimmbuffer}.
The DRAM read latency in the memory controller, $tRL$, can be increased by 6.8$ns$,
which is still within the adjustable range of commodity processors.
For a slightly more complex system with two layers of extending hardware and 
minimal logic processing, the propagation delay will likely approach 20$ns$, 
which is difficult for commodity processors to tolerate.

\subsection{Memory Extension over Custom Memory Interfaces}

There are many proposals for replacing the synchronous 
memory interface and breaking the tight processor-memory coupling. 
Typically, an interface buffer/die is introduced to bridge 
processor and memory, such as FBDIMM~\cite{fbdimm}, BOB~\cite{cooper2012buffer}, and HMC~\cite{hmc}.
Other researchers study more exotic organizations; for instance,
Chen et al.~\cite{chen2014mims} study a message-based memory subsystem with high-speed serial links;
Fang et al.~\cite{fang2011memory} incorporate emerging technologies on DDRx-like buses;
and Udipi et al.~\cite{udipi2011combining} look at using 3D stacking
and photonics to create scalable memory systems. 
The protocol between the buffer/die and memory is still a synchronous DRAM protocol,
but the protocol between the buffer/die and processor is replaced by a packet-based
access protocol.
Although such methods are effective, they require changes to the memory controller and processor-memory interface. 
It is uncertain whether processor vendors will accept such solutions. Even if they do, 
high cost and increased access latency may still limit 
adoption, e.g., BOB is only supported in high-end product lines.

\subsection{Memory Extension over Inter-Processor Interfaces }

A coherent network can connect multiple server processors to form
a NUMA (Non-Uniform Memory Access) node. Each processor can access memory on other nodes with
additional latency, which extends the total capacity of directly addressable memory. 
Various NUMA systems are available from 
low-end, dual-socket systems to those with 100s of CPUs~\cite{sgi}. 
When it comes to  memory capacity, though, NUMA is expensive. 
First, adding more memory modules necessitates adding more processors, 
which may be wasteful for memory-bound applications. Second, 
maintaining cache coherence across shared memory incurs significant overheads. Both complexity 
and cost are added to the processor, e.g., only high-end processors support NUMA 
with more than two processors. Third, the access latency across a NUMA interconnect is relatively 
high, e.g., the Intel Quick Path Interconnect (QPI) adds about 58-110$ns$ latency 
per hop~\cite{molka2009memory}.

\subsection{Memory Extension over Network Interfaces}

PCIe is an asynchronous, packet-based 
protocol. The lack of a latency constraint facilitates 
the design of scalable DRAM organizations over standard PCIe interface, 
but the latency of accessing memory via PCIe is several microseconds. 
Lim et al.~\cite{lim2009disaggregated,lim2012system} find that 
page-swapping between local and remote memory via DMA
performs reasonably for applications with high locality,
but swapping is inefficient for applications with large working sets and irregular access patterns.
Besides, data accessed via PCIe are not directly cacheable.
Other proposals 
can use memory on remote servers or memory blades
via a network interface. Software approaches like 
vSMP~\cite{vsmp} and MemX~\cite{deshpande2010memx}
access memory on remote servers over commodity InfiniBand or Ethernet. 
For all I/O interface based schemes, the latencies are difficult to go below one micro-second.

\subsection{Memory Extension with Emerging Technologies }

Emerging storage class memory (SCM) technologies such as PCM and ReRAM could 
provide higher storage density and lower power consumption than DRAM
at comparable access latencies, making it a potentially good  
candidate for capacity extension. However SCM 
has more timing constraints that are quite different with DRAMs. For example, 
write latencies are about 10$\times$ longer than read latencies, 
and reads are still 2-3$\times$ slower than DRAM~\cite{lee2009architecting}. 
So it is not possible to access SCM through the commodity 
DDRx SDRAM interfaces without modifying the processor-integrated memory 
controller. Micron recent PCM chip~\cite{micronpcm} uses a special 
JEDEC LPDDR2-N~\cite{lpddr2} interface that adds a PREACTIVE 
command and an overlay window especially designed for NVM. 
SanDisk's UlltraDIMM connects NAND flash to CPU through DDR3 Interface but only supports direct access to internal buffer~\cite{ulltradimm}.
Since SCM technology is still evolving, it is unlikely that 
a universal interface will be well defined and adopted by processor community soon.

\section{Overview}
\label{section:overview}

Both the DDRx and PCIe are widely-used, open standards, which makes their
interfaces good candidates for memory extension. We choose the memory interface
because of two advantages: latency and concurrency.
Even with the extra propagation delay, the access latency to 
extended memory is still within tens of nanoseconds.
And since these accesses are cacheable,
non-blocking caches help mask delays. 

\begin{figure}[!t]
\centering
\includegraphics[width=8.5cm]{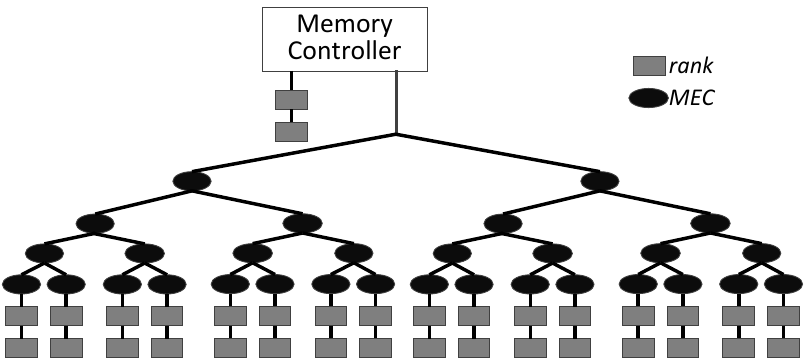}
\caption{Four-layer tree topology, assuming that a channel can 
only drive dual ranks at high frequency~\cite{rdimmcnt}}
\label{figure:topology}
\end{figure}


Figure~\ref{figure:topology} illustrates one potential memory extension
topology that allows us to populate more than two ranks per channel.
The Memory Extending Chips implement one slave and one master
DDRx interface. The slave of the \emph{top-level MEC} (MEC1) connects to the processor's
commodity memory interface, and the master connects to the DIMMs
or to the slave interfaces of other MECs.
The MEC interfaces are similar to those of LRDIMM~\cite{cisco}, but 
the internal logic and associated software break the one-layer constraint.



Since DRAM reads to extended memory experience intolerable
round-trip times, we must access the data in a way 
that breaks the $tRL$ constraint.
Note that both loads and stores cause DRAM reads. Stores first trigger
read-for-ownership (RFO) operations to bring data into cache.
They update the data in cache, and on eviction they write back to memory. 
We thus first discuss how loads work before discussing stores. 

\subsection{Load Operations}
\label{section:load}

To break the latency constraint, we could first prefetch the 
target data into the buffer of MEC1
and insert a line of fake data (e.g., repetitive patterns of 0x5a) as a 
placeholder in the processor's caches. A second demand load could then fetch
the real data from the MEC.
This scheme presents three challenges.
First, we can neither issue two normal loads
to the target address nor use a software prefetch instruction,
since those would cause the demand load to hit in cache and load the 
placeholder data. The demand load must reach
MEC1 to get correct data. Second, MEC1 must process the prefetch first. 
Modern processors typically employ out-of-order (OoO) 
execution in the instruction 
pipeline or the memory controller queue, which makes the order in which 
the loads reach the MEC unpredictable. 
Third, the MEC must issue the prefetch early enough to guarantee that the 
data will be loaded into its buffer before the demand load fetches the data.




\begin{figure}[!t]
\centering
\includegraphics[width=7.5cm]{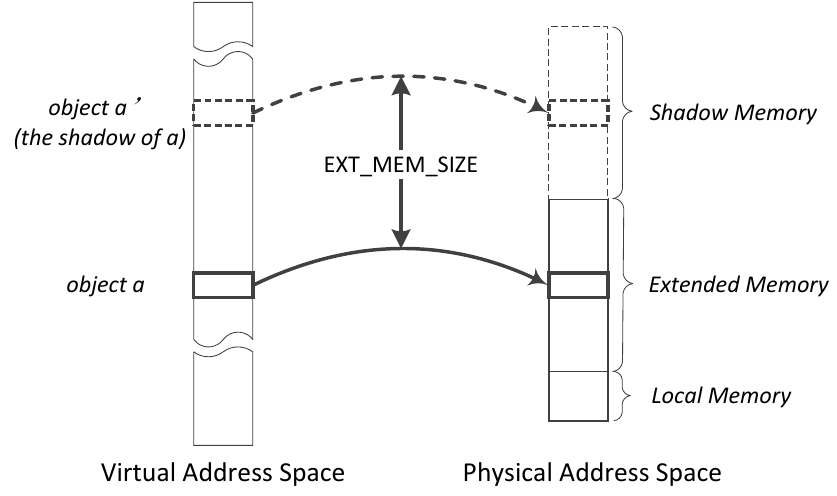}
\caption{Relationship between virtual and physical memory spaces. The shadow 
space does not map to real DRAM. 
}
\label{figure:addrspace}
\end{figure}

To address the first challenge, we manipulate the data addresses so
that the processor thinks the prefetching load is to a different location
from the demand load, but the MEC knows that they correspond to the
same target location. For instance, adding a flag bit suffices to distinguish
the addresses inside the processor. The MEC simply ignores this bit.
This creates a ``shadow'' address for each location in the extended memory.
Figure~\ref{figure:addrspace} shows the relationships among the local, extended,
and shadow memory spaces.

To address the second and third challenges, we design two different twin-load mechanisms.
We also investigated an extreme way to guarantee strong
ordering by making the extended memory uncacheable~\cite{intelsdm},
but we omit those results here due to its poor performance. 

The first mechanism, \textbf{TL-LF}, inserts a \emph{load fence} instruction
between the loads. The fence guarantees that the second load 
will not execute until the first receives data from memory~\cite{intelsdm}. 
The spacing between the loads is the round-trip time of a memory load,
which is large enough to tolerate the propagation delay.
This mechanism retains the benefit of cache locality, but blocks 
all loads following the fence.

To better exploit memory concurrency, we need to allow more out-of-order execution.
Our second solution, \textbf{TL-OoO}, does not designate which is the prefetch or demand load
at the software level, but assigns appropriate identities dynamically
via both hardware and software. The load that arrives first triggers
the prefetch and returns fake data, and the one that follows returns the true data.
Software identifies which data is correct on-the-fly.

If both the extended and shadow addresses map to the same DRAM bank,
the second twin-load artificially triggers a DRAM row
miss that forces TL-OoO to delay the load's MEC arrival.
Recall that an RD command to the same bank but 
different row must wait $tRTP$ time before 
issuing the PRE to close the current row, $tRP$ time to complete the PRE and 
issue the ACT for the new row, and finally $tRCD$ time to complete the ACT and 
issue the RD.
The minimum total delay is about 35ns at DDR3-1600, 
which is enough to tolerate propagation
delays for up to five MEC layers.

\subsection{Store Operations}
\label{section:store}

We need twin-loads to bring data into cache before we perform store operations.
Memory consistency requires us to ensure that data always be written to 
the true cache line. 
In the unlikely event that an interrupt happens between the twin-load and the store, 
the correct cache line may have been evicted by the time the store is resumed. 
The store has to trigger an RFO operation to load a fake line into cache, 
the modification of which will cause an error.
To avoid this, an atomic \emph{compare\_and\_swap}
(CAS) instruction\footnote{CAS is named CMPXCHG in the x86 instruction set~\cite{intelsdm}.}
first compares the correct value obtained from the twin-load with the value in the 
cache line, and swaps (stores) the new value into the cache line only when the comparison 
succeeds. Then if the RFO after the interrupt brings the 
fake data into cache, the comparison will fail, the cache line remains unmodified,
and the store is retried. 

\section{Implementation Details}
\label{section:details}




The MECs organize the physical DIMMs/ranks/banks into logical DIMMs/ranks/banks
that the memory controller sees. 
MEC1 asserts a fake serial presense detect (SPD)~\cite{spd} to the memory
interface, and all MECs maintain simple mapping tables.
MEC1 chooses one address bit to differentiate the extended and shadow memories.
For TL-OoO, that bit must be a row address bit, and since
memory controllers generally use the most significant bit (MSB) of the 
physical address in the row address, we choose it.
For simplicity, TL-LF also uses the MSB, even though it affords more
flexibility in how memory capacity grows.
The physical memory space consists of 
the local memory, extended memory, and shadow memory. 
Figure~\ref{figure:addrspace} shows
that only local memory and extended memory physically store data. 





\label{section:offset}

\subsection{Software Modifications}

\begin{figure}[!t]
\centering
\includegraphics[width=8.5cm]{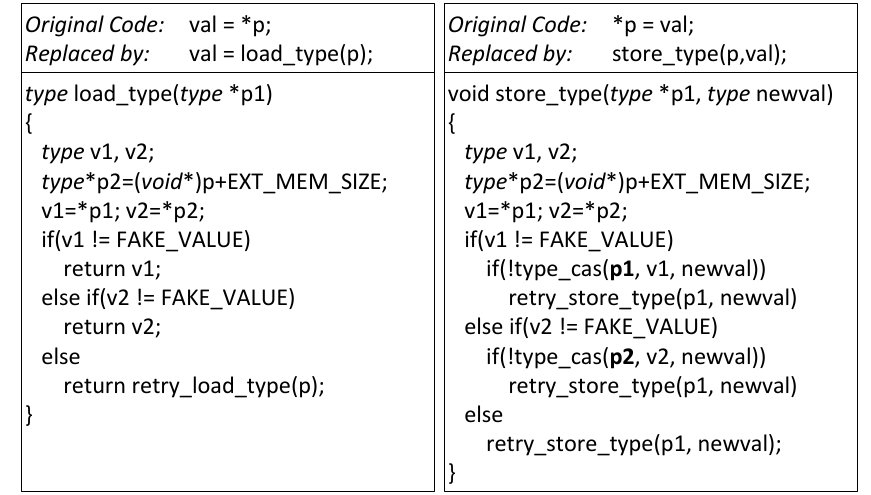}
\caption{Semantics of twin-loads in TL-OoO}
\label{figure:macro}
\end{figure}



The programmer must identify which objects to place
in extended memory. Large data objects make good candidates, whereas
the OS, code, stack, and small
objects should be in local memory.
The programmer use a special interface 
to allocate objects in extended memory.
For the most complicated application we evaluate, 
the modifications took less than two days (including time to understand the code).
For applications that index large arrays to access data,
they took less than half a day.

Figure~\ref{figure:macro} shows how loads and stores to identified objects in the program are
replaced by (inlined) functions that implement TL-OoO.
Loads to virtual address \emph{p} 
are replaced by the function \emph{load\_type(p)}, which loads both $p$
and $p'$ concurrently, and compares their return values to identify the correct one. 
Stores to virtual address \emph{p} are
replaced by the function \emph{store\_type(p,val)}.
Two functions, \emph{retry\_load\_type(p)} and \emph{retry\_store\_type(p,val)}, 
handle the cases in which both loads return fake values or the atomic
CAS fails. 

Such modifications can be done automaticlly by a compiler with user-annotations. This work will be introduced in our future paper.

\subsection{Extended Memory Management}
\label{section:allocator}

%
Modifying the OS is a practical means of managing the
extended memory, but for this study we choose to implement 
a lightweight manager outside the OS. 
Big memory applications usually allocate most memory during initialization,
with few changes to the allocations throughout execution~\cite{basu2013efficient,reiss2012heterogeneity}.
For example, Memcached~\cite{memcached} preallocates a big chunk of memory  
at startup and self-manages it to allocate items internally.
Such applications need no
complex managers to minimize fragmentation due to frequent
allocations and deallocations.

The extended and shadow memory spaces are reserved by the OS at boot time 
and can be allocated in user space via \emph{mmap()}. 
To simplify memory management, we allocate/deallocate extended and shadow 
memory together in large blocks (e.g., 64MB).
When allocating a block, two virtual memory regions at a distance of
EXT\_MEM\_SIZE are allocated, as well. Both the virtual
and  block physical addresses are passed to two \emph{mmap()}
calls to construct corresponding virtual-to-physical mappings in the page table.
If the virtual address of an object in extended memory is \emph{p},
the corresponding virtual address for its shadow \emph{p'} 
is simply \emph{p}+EXT\_MEM\_SIZE, as shown in Figure~\ref{figure:addrspace}.

\subsection{Twin-Load Processing}

\begin{figure}[!t]
\centering
\includegraphics[width=8.5cm]{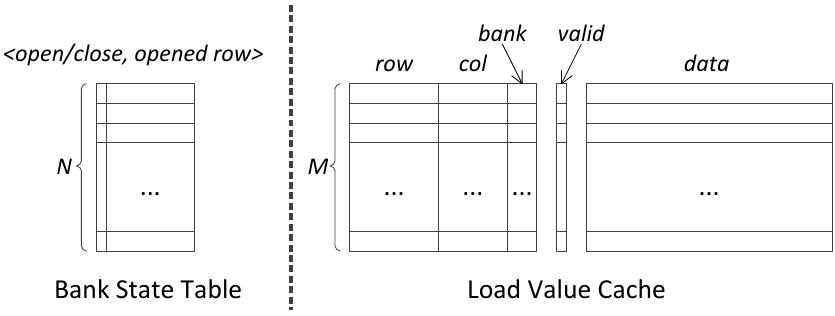}
\caption{The Bank State Table and Load Value Cache in the MEC1 hardware. 
$N$ and $M$ are the number of entries, respectively, where $N$ is
equal to the number of logical banks, and
$M$ is a design parameter.
}
\label{figure:table}
\end{figure}


In a multi-layer extended system, MEC1
identifies the two loads and forwards
the request on the first load, temporarily buffers the data, and returns it on the 
second load. The MECs on the other layers are much simpler,
either executing the received commands or forwarding them to the next layer. 

MEC1 maintains two main structures not required in lower MECs:
the Bank State Table (BST) and the Load Value Cache (LVC), illustrated in 
Figure~\ref{figure:table}. The routing table required to implement
tree topologies is not shown. 
For each logical bank, a BST entry indicates whether the 
bank is open and stores the address of the last row opened.
The LVC entries temporarily store prefetched values 
for the first loads. The tag is the address, and the valid
bit indicates whether the entry is in use.
Our LVC uses LRU replacement.


Upon receiving an ACT command, 
MEC1 records the bank's row address in the BST. When the 
RD arrives with its bank address, the MEC accesses the BST 
to retrieve the previously issued row address. 
The load address is reconstructed as <row, column, bank>. 

For TL-OoO, 
the address is the tag for the LVC lookup: if the lookup misses, 
the access should be the first load, otherwise the second load.
When MEC1 sees the the first load, 
it allocates an LVC entry, sets the tag 
to the load address, sets the valid bit, and forwards the RD 
to the target MEC. After $tRL$ time, it puts the fake values
on the data bus to the memory controller. The target MEC fetches
the data from DRAM and returns it to MEC1 with the LVC entry ID,
where it is inserted into the MEC1 LVC. 

In general, the LVC size $M$ should try to guarantee that 
the corresponding entry will not have been evicted when the data return, i.e.,
$M>(2\times tPD+tRL)/tCCD$,
where $2\times tPD+tRL$ is the round-trip time for returning data, and $tCCD$ 
defines the minimum interval between consecutive RDs. 
For TL-OoO, the maximum tolerable propagation 
delay is 35ns, and thus $M>10$ suffices.
TL-LF can tolerate much longer delays, and so
the LVC must be larger.
When MEC1 identifies the RD as the second load, it returns the data
to the memory controller after $tRL$ time. The valid bit is cleared 
to free the entry. 

If the second load arrives too late, the data may have been
evicted from the  MEC1 LVC.
For TL-LF, the second load then returns the fake value.
For TL-OoO, the MEC will identify the intended second load as the first, 
reallocate a new cache entry, return fake values, and prefetch the data again. 
Software retry ensures that the load gets correct data, but
we want to avoid such cases, which waste prefetches and hurt performance. 
By monitoring 
our prototype's DRAM command bus, we find that twinned loads are 
separated by an average of six other loads, which can guide the design choice of $M$.

The middle MECs forward commands to the target leaf MECs
to execute. They use the high bits of the 
row address as the physical DIMM ID for command forwarding. 
The routing table determines the forwarding port. For the ACT command, 
the ID is in the row address. For other commands, 
MEC1 gets the ID from the BST and passes it with the command.

\subsection{Cache State and Correctness}

\begin{table}[!tb]
\centering
{
  \caption{Twin-Load Results with Respect to Cache State}
\sffamily
\footnotesize
{
  \begin{threeparttable}
  \begin{tabular}{|c|c|c|c|c|}
    \hline
    State & $v$ & $v'$ & DRAM Reads & Result\\
    \hline
    \hline
    1 & not in cache & not in cache & two & $v,v'$\\
    \hline
    2 & in cache & in cache & zero & $v,v'$\\
    \hline
    3 & in cache & not in cache & one & $v,v'$\\
    \hline
    4 & not in cache & in cache & one & $\bm{v',v'}$\\
    \hline
  \end{tabular}
{
}
  \end{threeparttable}
  \label{table:cache}
}
}
\end{table}
 
Since both extended and shadow memory are cacheable, one or both
of the twin-load accesses might not reach MEC1. 
Let $v$ be the correct value and $v'$ the fake value.
Table~\ref{table:cache} lists their possible states with respect to the cache.
In State 1, the initial state, both loads trigger 
DRAM reads. The MEC takes the first 
read as the prefetch and returns the fake value and then returns the 
correct data on the second read. 
In State 2, the MECs are not involved. Both 
loads commit quickly, one with the correct value and one 
with the fake value.
In State 3, one load returns the correct value directly 
from cache, and the MEC identifies the other load as the prefetch 
and returns the fake value. The corresponding LVC entry will eventually be evicted.
In State 4, one load hits in 
cache and returns the fake value, and the other causes a DRAM
read that \emph{also} returns the fake value, since no former prefetch has
reached the MEC.

We fall back to a software retry to handle this case: both
load addresses are first invalided to return to state 1, and then we use 
another twin-load to get the correct data. A memory fence
instruction is required to complete the invalidation before the following twin-load.
If the retry also gets the fake value, we throw an exception that
invokes a safe path to memory. 
We are investigating other strategies for better performance, but discussing
them is beyond the scope of this paper.

\subsection{Exception Handling}

There are two rare cases in which the retry may fail: the LVC entry gets evicted
before the second load arrives, or the correct data is the same as the fake
value. Our solution is to implement a slow but safe path by which to load
the data. We add three uncacheable memory mapped 
registers in MEC1: an address register to receive the 
physical load address, a flag register to indicate load 
completion, and a data register to hold the loaded data.
The exception handler actions are like reading I/O ports.


\section{Evaluation Methodology}
\label{section:methodology}

\begin{table}[!b]
\centering
{
  \caption{Emulated Systems}
\sffamily
\footnotesize
{
  \begin{threeparttable}
  \begin{tabular}{|l|c|c|c|c|}
    \hline
    \textbf{System}	& TL		& NUMA		& PCIe		& Ideal		\\
    \hline
    \hline
    Processor		& \multicolumn{4}{c|}{One Intel Xeon E5-2640 Processors (6-core, 12-thread)~\tnote{1}}	\\
    \hline
    Local Memory	& \multicolumn{2}{c|}{0-8GB}	& 0-$x$GB\tnote{2}	& 0-32GB	\\
    \hline
    Extended Memory     & \multicolumn{2}{c|}{8-32GB}	& $x$-32GB\tnote{2}	& -		\\
    \hline
    Shadow Memory       & 40-64GB	& \multicolumn{3}{c|}{-}	 		\\
    \hline
    Extended Interface	& DDRx 		& QPI		& PCIe		& - 		\\
    \hline
    Access Mechanism	& TL-LF/TL-OoO	& cc-NUMA	& Swapping	& Ideal		\\ 
    \hline
  \end{tabular}
{
  \begin{tablenotes}
    \scriptsize
    \item[1] We have two processors, but use only one for program execution for all systems.
    \item[2] We vary the cut-off point to emulate different local:swapped ratios.
  \end{tablenotes}
}
  \end{threeparttable}
  \label{table:systems}
}
}
\end{table}

\begin{table*}[!tb]
\centering
{
  \caption{Workloads}
\sffamily
\footnotesize
{
  \begin{tabular}{|c|c|c|c|c|}
    \hline
    Benchmark   &	Source			&	Type				&	Description		& Proportion in extended memory \\
    \hline
    \hline
    GUPS      	& HPC Challenge~\cite{hpcc}	& Micro-Benchmark			& Random access			& 100.00\%	\\
    \hline
    Radix	& PARSEC3.0~\cite{parsec}	& Kernel				& Integer sort			& 100.00\%	\\
    \hline
    CG		& NPB2.3~\cite{npb}		& \multirow{2}{*}{Scientific Computing}	& Calculating conjugate gradient	& 99.43\%	\\
    \cline{1-2} \cline{4-5}
    FMM		& PARSEC3.0~\cite{parsec}	& 					& N-body simulation		& 94.39\%	\\
    \hline
    BFS		& Graph500~\cite{graph500}	& \multirow{3}{*}{Graph Application}	& Breadth-first search		& 99.79\%	\\
    \cline{1-2} \cline{4-5}
    BC		& SSCA2.2~\cite{ssca}		&					& Calculating connection centrality & 76.92\%	\\
    \cline{1-2} \cline{4-5}
    PageRank	& In-house implementation	& 					& Calculating website ranks~\cite{brin1998anatomy} & 87.93\%	\\
    \hline
    ScalParC    & NU-MineBench~\cite{minebench}	& \multirow{2}{*}{Data Mining}		& Parallel classification	& 94.48\%	\\
    \cline{1-2} \cline{4-5}
    StreamCluster& PARSEC3.0~\cite{parsec}	& 					& Online clustering		& 92.93\%	\\
    \hline
    Memcached	& Memcached-1.4.20~\cite{memcached} & Data Serving			& Key-value caching system	& 97.30\%	\\
    \hline
  \end{tabular}
{
}
  \label{table:workloads}
}
}
\end{table*}

To evaluate our twin-load implementation, we emulate multiple sytems for comparison:
\begin{itemize}
\item TL-LF and TL-OoO: our twin-load mechanisms with local, extended memory, and shadow memory;
\item NUMA: a system using QPI to connect more processors so they can attach more memory;
\item PCIe: a system using PCIe to connect more memory, which is accessed using page swapping~\cite{lim2009disaggregated,lim2012system}; and
\item Ideal: an ideal system with all memory locally attached.
\end{itemize}

\noindent
Our host system has two processors and eight 8GB DIMMs (64GB in total). 
Table~\ref{table:systems} shows how the host memory is used to emulate 
the extended memory systems.
For the TL, PCIe, and Ideal systems, we attach all DIMMs to a single processor
(and execute only on that processor)
to avoid performance variations among different runs due to nondeterministic 
memory-to-processor affinities. Our experiments are independent
of any specific topology --- as long as the propagation delay in within 35ns,
the software behaves the same.

For the TL systems,  
both the extended and shadow memories are emulated using reserved host memory.
The shadow memory is initialized to hold fake values to emulate the MEC functionality.
The twinned addresses cause DRAM row misses in the host memory controller.
We implement the required software to generate twin-loads to the extended and shadow memories
for certain accesses.
For the NUMA systems, 
we attach one DIMM to one processor to emulate local memory and attach seven DIMMs to the other 
processor to emulate extended memory. Programs execute on the former 
and access extended memory via QPI. 
For the PCIe system,
we emulate extended memory as a host-memory RAM Disk configured 
as a swap partition. We emulate the remote page swapping with default Linux 
swapping to the RAM Disk.
For the Ideal system, we emulate
all local memory as host memory. No software change is required.

Note that there are some deviations between a real twin-load system and the emulated
system. In the emulated system:

\begin{enumerate}
\item
loads to the extended and shadow memories always return the correct and 
fake values, respectively, thus it is possible that the correct value returns earlier 
and advances program prematurely; and
\item
for the fourth case in Table~\ref{table:cache}, the missed load always returns the 
correct value from memory, whereas it should 
return the fake value and trigger a retry.
\end{enumerate}
To avoid the first situation, we choose to advance the program only when both the 
values have been returned and checked. This is a conservative choice, since for the first 
and third cases in Table~\ref{table:cache}, it could be that the correct 
value in cache is compared first, and the program could proceed without waiting for 
the result of another load. 
It is difficult to avoid the second situation because the software 
cannot know the cache state. However, by recording the memory requests on the
memory bus using a tool like a DDR3 protocol analyzer~\cite{lecroy}, we find that over
96\% of loads to extended memory are twinned. This can be easily explained, since the two addresses
are always accessed synchronously and are very likely to be brought into and evicted from
cache together. Taking into account the conservative policy for the first deviation, we believe
our emulation reasonably approximates a real extended system.



Table~\ref{table:workloads} lists the workloads we use in our evaluations. 
From a variety of application domains, we select 10 benchmarks
with footprints that scale easily and code sizes that are reasonable for manual modification.
For Memcached, a client running \emph{memslap}~\cite{memslap} is connected to the Memcached server 
via Gigabit Ethernet; to avoid the network bandwidth becoming bottleneck, we test 
small objects~\cite{lim2013thin}
and only use four threads on the server side.

For all benchmarks, we evaluate two footprints --- a medium one around
4GB and a large one around 16GB. For the TL and NUMA systems, we modify source code 
to allocate large objects in extended
memory. Table~\ref{table:workloads} shows the proportion of data in extended memory. 
For the PCIe system, we let the Linux swap mechanism manage data placement. 
We use performance counters to gather architectural statistics.

\section{Results}
\label{section:results}

\begin{figure*}[!t]
\centering
\includegraphics[width=16cm]{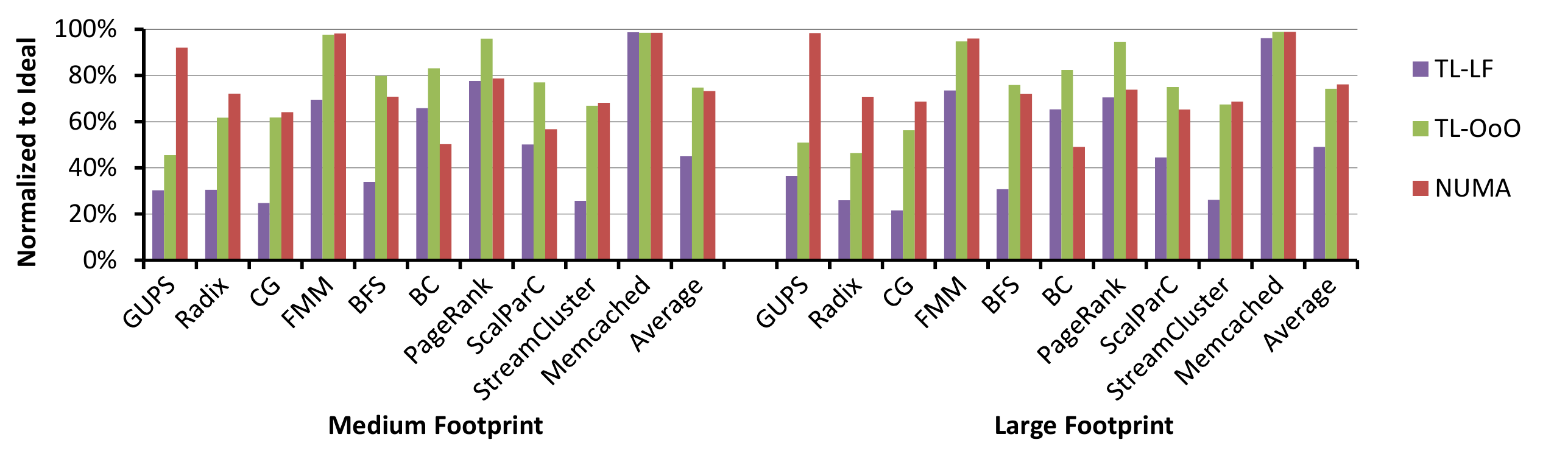}
\caption{Normalized Performance of Various Mechanisms}
\label{figure:perf}
\end{figure*}
  
We compare TL-LF and TL-OoO against the NUMA and
Ideal systems described in Section~\ref{section:methodology}.
Figure~\ref{figure:perf} shows experimental results for these mechanisms
on our emulated prototypes. We normalize performance relative to Ideal.
TL-LF, TL-OoO, and TL-NUMA achieve 45\%, 75\%, and 73\% of Ideal performance
for medium footprints, and 49\%, 74\%, and 76\% of Ideal performance for large
footprints.
This suggests that footprint size does not significantly affect performance, 
so we restrict our discussion to large-footprint results.


\subsection{TL vs. Ideal}

We first discuss the potential penalties for twin-load compared
to having all local memory. Then we 
discuss how TL-OoO might alleviate certain penalties.
Finally, we discuss the shortcomings of TL-LF and discuss 
future optimizations.

\paragraph{Potential Penalties.}

\begin{figure}[!t]
\centering
\includegraphics[height=3.5cm]{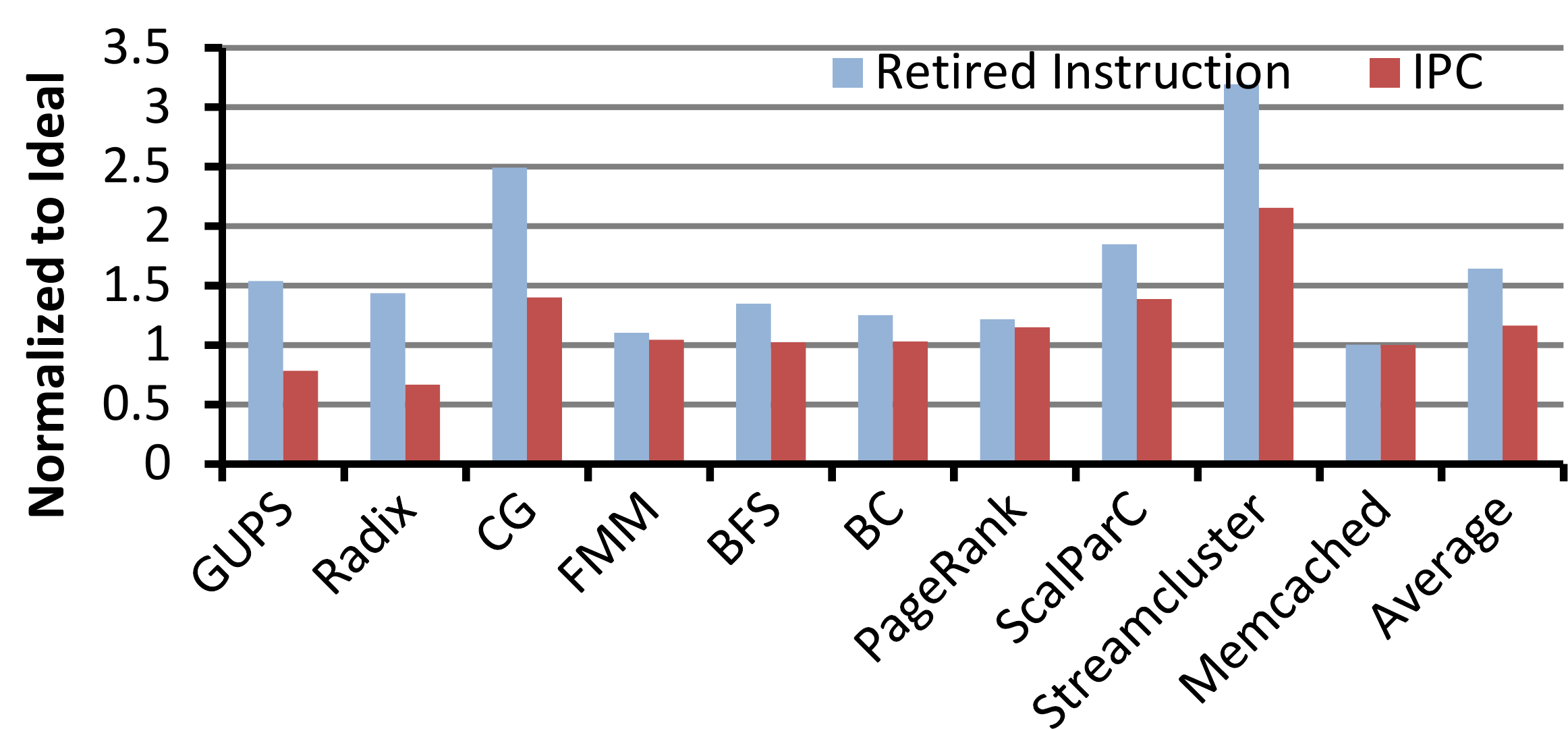}
\caption{Instruction Count and IPC of TL-OoO Relative to Ideal}
\label{figure:inst-ipc}
\end{figure}
  
\begin{figure}[!t]
\centering
\includegraphics[height=3.5cm]{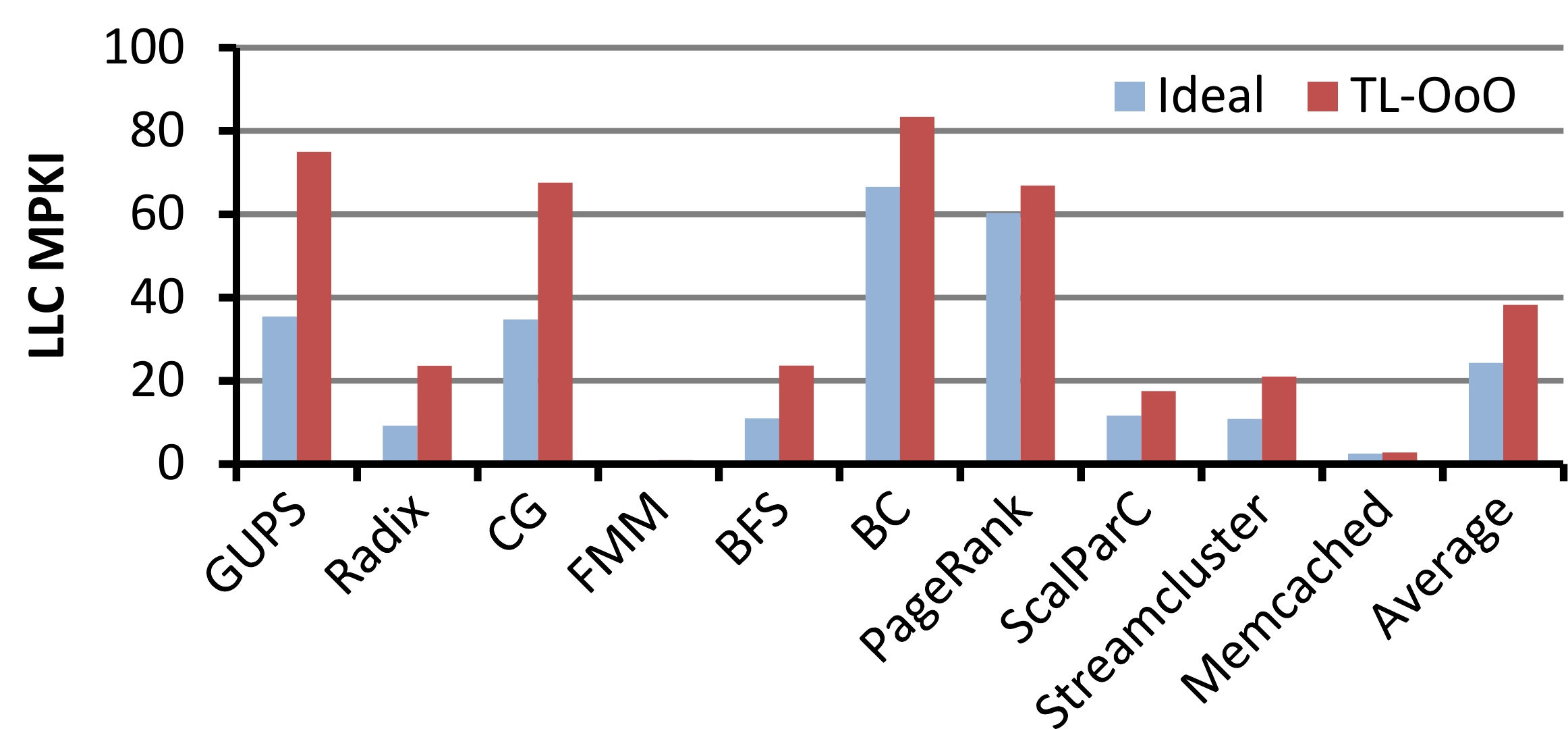}
\caption{LLC MPKI}
\label{figure:llc-mpki}
\end{figure}
  
\begin{figure}[!t]
\centering
\includegraphics[height=3.5cm]{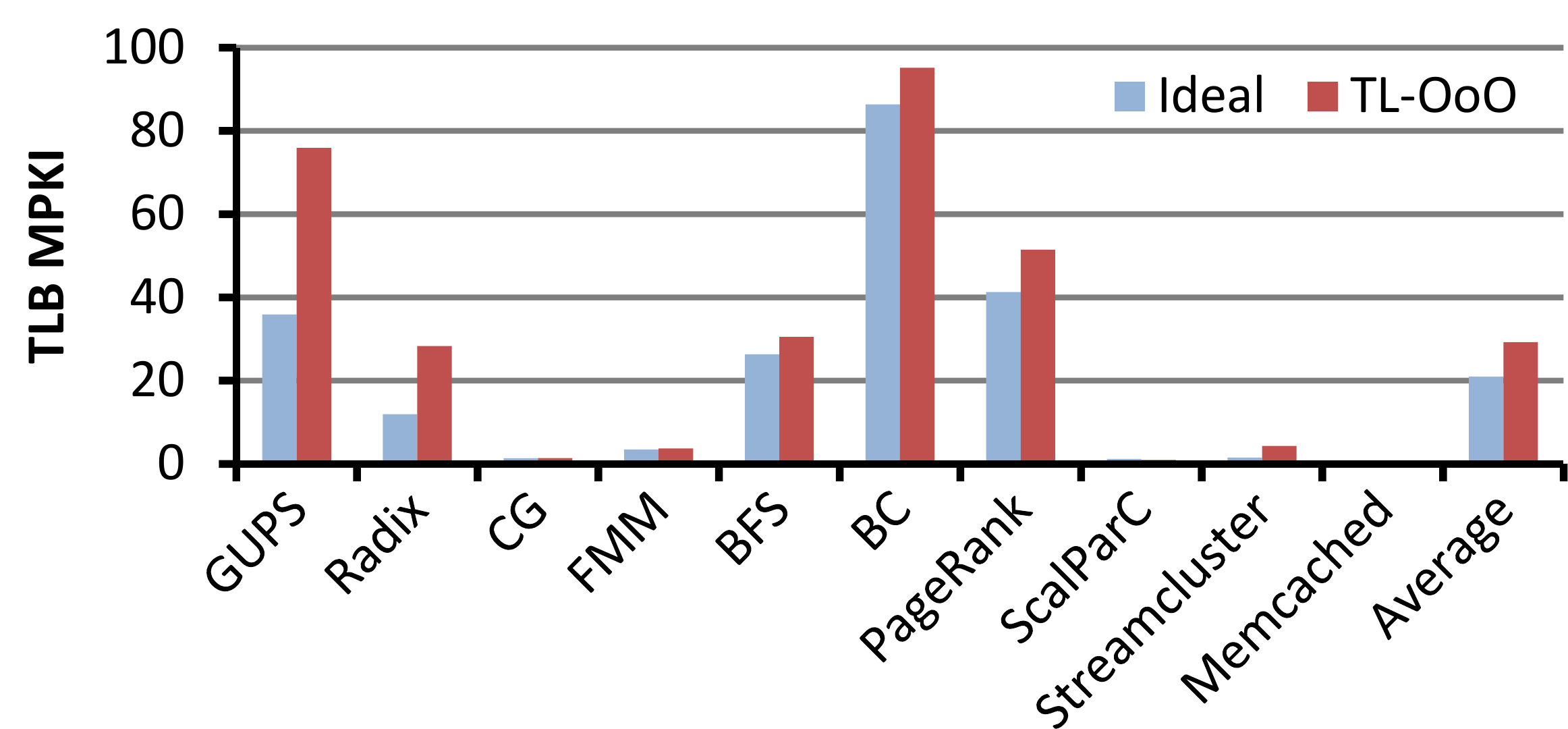}
\caption{TLB MPKI}
\label{figure:tlb-mpki}
\end{figure}

Twin-load obviously increases the number of instructions and data 
accesses, potentially causing more cache misses. Also, since we double the address space,
TLB conflicts will be exacerbated. 
Figure~\ref{figure:inst-ipc}, Figure~\ref{figure:llc-mpki}, and Figure~\ref{figure:tlb-mpki}
show the effects of twin-load on instruction execution, LLC, and TLB behaviors.
(Note that the LLC MPKI and TLB MPKI of TL-OoO are relative to the number of retired 
instructions in the Ideal case, to show the absolute miss increase.)
Since twin-load replaces some load/store instructions with inline functions,
the increas in number of instructions retired depends on the proportion of memory accesses
and their relative proportion targeting extended memory. 
TL-OoO's retired instruction count increases by 64\%, on average.
LLC misses increase by 11-156\% (71\%, on average). 
If all data are in extended memory, the number of LLC misses can potentially
double, as is the case for GUPS, Radix, CG, and BFS.
For others (e.g., BC, PageRank, and ScalParC) a small portion 
of local data may contribute to a significant portion of the accesses, thus
twin-load only modestly increases LLC misses.

Figure~\ref{figure:tlb-mpki} shows that
workloads with significant TLB conflicts can be classified into two categories: 
graph applications and applications that store most data structures in extended memory.
For instance, doubling the extended address space roughly doubles the TLB misses
for GUPS and Radix.
For the graph applications, our results suggest that the relative small but frequently 
accessed vertex-associated metadata (rather than the large graph) contribute to 
most of the TLB misses. This is because such metadata are randomly accessed and
large enough to exceed the TLB coverage (2MB for a 512-entry TLB with 4KB pages),
and the graph traversal thrashes the TLB.
Workloads with relatively good locality (e.g., CG and ScalParC) 
cause few TLB conflicts.
Increases in TLB MPKI range from 3\% to 179\% (39\% on average).

\paragraph{Potential Benefits for TL-OoO.}

\begin{figure}[!t]
\centering
\includegraphics[height=3.5cm]{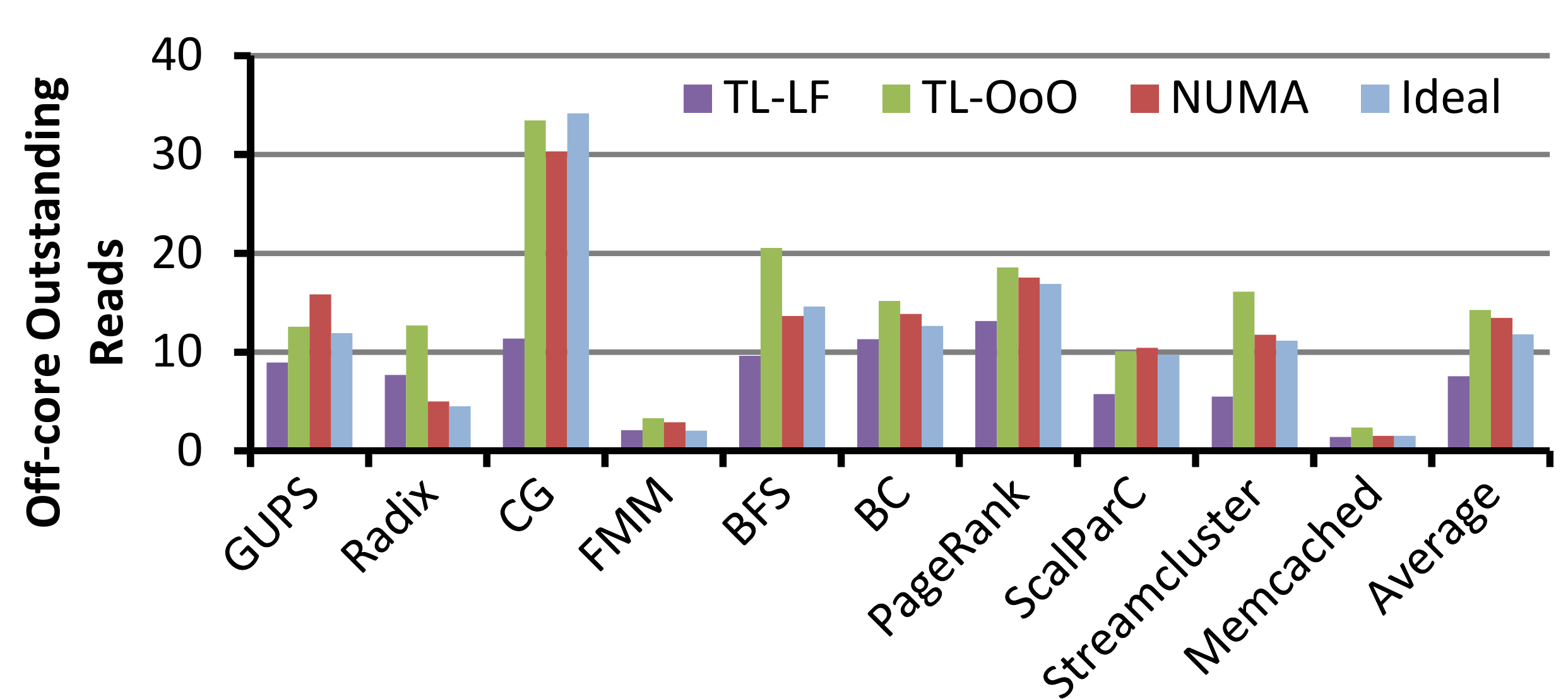}
\caption{Average Number of Off-Core Outstanding Reads}
\label{figure:offcore-read}
\end{figure}
 
\begin{figure}[!t]
\centering
\includegraphics[height=3.5cm]{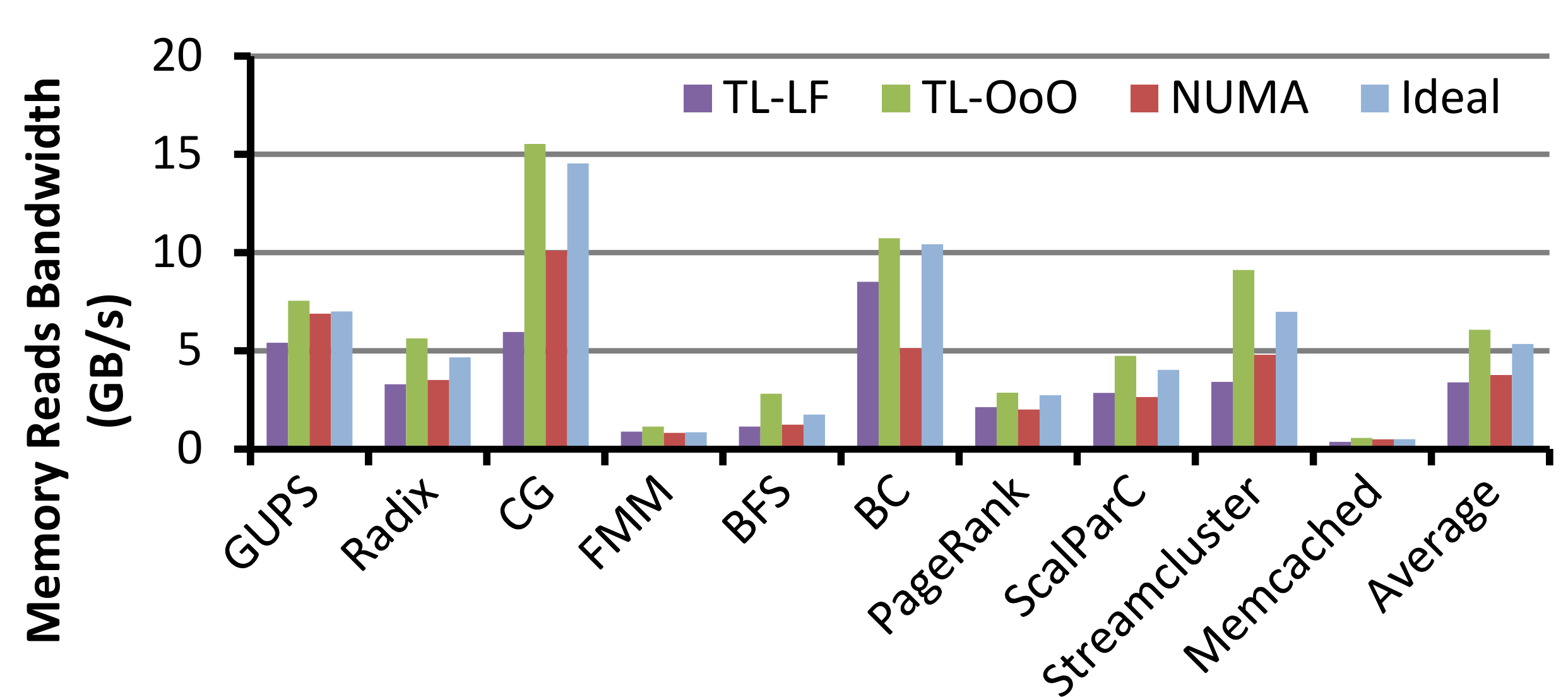}
\caption{Average Read Bandwidth}
\label{figure:read-bw}
\end{figure}
 
Although twin-load increases the number of executed instructions by 64\% and LLC misses 
(long latency memory accesses) by 71\% compared to Ideal, average performance slowdown
for TL-OoO is only 25\% and 26\% for medium and large footprints, respectively.
This is due to twin-load's better processor utilization:
Figure~\ref{figure:inst-ipc} shows that even though twin-load increases the number of instructions,
it delivers higher IPCs for most workloads. 
(The remaining gap reflects the performance slowdown.)
The pipeline usually stalls on long-latency memory accesses, but our
twin-load instructions can exploit such stall slots, masking the increase
in non-memory instructions.

Achievable memory-level parallelism (MLP) of most applications is limited,
which is far from saturating the processor's available memory access 
concurrency, which is defined by the number of MSHRs. 
Thus TL-OoO can take advantage of the remaining capacity for concurrency 
to overlap execution of the extra loads.
Figure~\ref{figure:offcore-read} shows that the average number of  
outstanding off-core reads increases from 11.8 to 14.3, especially
for those workloads with significant increases in LLC misses (except GUPS and CG). GUPS's 
concurrency is likely limited by the many TLB misses, while CG 
seems to saturate the hardware support for concurrency. 
Since we increase memory concurrency, the achievable memory bandwidth also
increases, as shown in Figure~\ref{figure:read-bw}. 

\paragraph{Shortcomings of TL-LF.}

The most obvious shortcoming of TL-LF is its limited memory concurrency. 
Consecutive accesses to extended memory are serialized 
by the load fence.
Figure~\ref{figure:offcore-read} and Figure~\ref{figure:read-bw}
show that the number of outstanding off-core reads and the memory read bandwidth
are both decreased by 34\%.
Although it incurs more than 50\% slowdowns, TL-LF can potentially 
tolerate higher latencies than TL-OoO, making it adaptable to more application cases.
A possible optimization for TL-LF is not to insert a fence per data access, but
to batch the first twin-load instructions for several accesses, insert the fence, and then 
perform the second twin-loads and software checks. We leave this for future work.

\subsection{TL vs. NUMA}

Figure~\ref{figure:perf} shows that TL-OoO exhibits comparable performance to NUMA.
On our host system, the access latency to local memory and remote memory (via QPI) is 
about 100$ns$ and 170$ns$, respectively. 
For NUMA, the 
long latency to extended memory and the limited memory concurrency cause 
memory throughput (bandwidth) to decrease by an average of 30\% (and up to 51\%) compared 
to Ideal, resulting in an average 24\% (and up to 51\%) performance slowdown.
TL-OoO performs much better than NUMA for graph applications, in particular.
The irregular access behaviors and corresponding limited intra-thread memory concurrency 
make graph applications latency-sensitive. Compared to the 70$ns$ 
latency increase for NUMA, TL-OoO incurs less relative penalty --- recall that a row miss
causes only 35$ns$ extra latency.

\subsection{TL vs. PCIe}

\begin{figure}[!t]
\centering
\includegraphics[height=3.5cm]{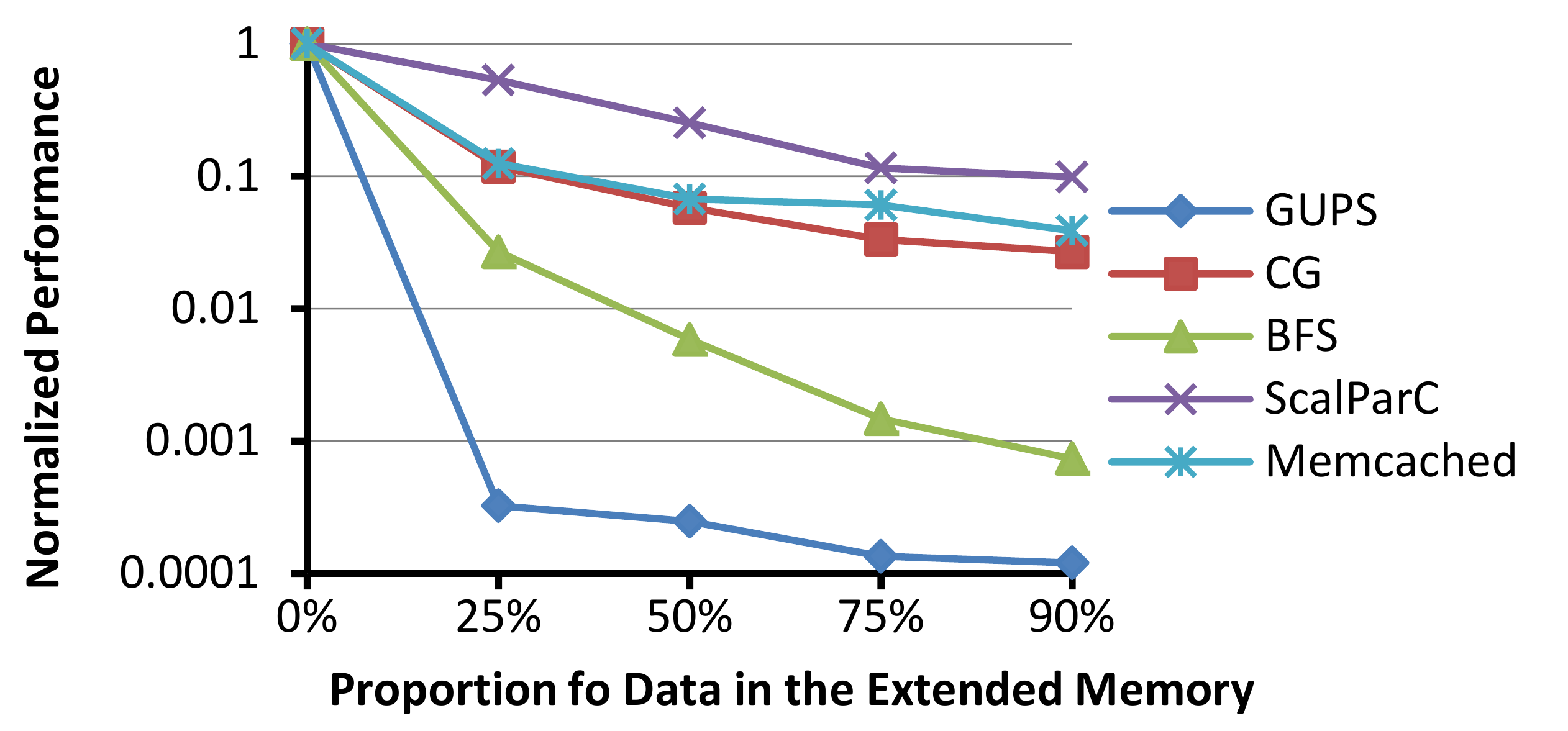}
\caption{Performance of PCIe Swapping mechanism}
\label{figure:perf-pcie}
\end{figure}

We evaluate the performance slowdown of using PCIe extended memory paging. In the experiment,
we change the ratio of data to be placed on extended memory, ranging from 0\% to 90\%.
The original replace procedure in Linux is designed for swapping with much slower hard disk so 
that it may be quite complicated and slow for swapping with fast PCIe remote memory.  
It takes about 7.8us to swap a page on our prototype, which is 1.4$\times$ of the fastest PCIe 
replacement policy~\cite{lim2012system}. Thus, to compensate for such extra software overhead, we double the 
measured performance of the emulated PCIe system in the comparison result. We choose five representative benchmarks
--- GUPS, CG, BFS, ScaleParC, and Memcached --- from 
Table~\ref{table:workloads}. 

Figure~\ref{figure:perf-pcie} shows results normalized to those of a non-swapping
(0\% data in extended memory) system. 
Placing only 25\% of data into extended memory slows performance
of most workloads by quite much. Performance further degrades as we move more
data into extended memory. At 90\%, swapping pages with PCIe extended memory 
yields slowdowns of one to four orders of magnitude.

When 25\% of data reside in extended memory,
ScalParC has the best performance (0.53$\times$), mainly due to its low LLC and TLB MPKIs, which
suggest good locality and less pressure on the memory system. 
The extremely random-access  GUPS only achieves 0.0003$\times$ the performance.
Even for Memcached, which shows insensitivity to the memory system (Figure~\ref{figure:perf}),
performance is only 0.13$\times$.
For CG and BFS, resulting performance is 0.12$\times$ and 0.27$\times$, respectively.
TL-LF and TL-OoO perform much better than this, even if we put over 90\% of data into 
extended memory.

\section{Discussion}

We recognize that accurate cost projections can be difficult,
we nonetheless try to put relative costs in perspective
before examining the impact of extending the DRAM $tRL$
time.

\subsection{Cost Analysis}

\begin{table}[!b]
\centering
{
  \caption{Costs of various memory extension mechanisms}
\sffamily
\footnotesize
{
  \begin{threeparttable}
\newcommand{\tabincell}[2]{\begin{tabular}{@{}#1@{}}#2\end{tabular}}
  \begin{tabular}{|c|c|c|c|c|}
    \hline
    \tabincell{c}{Costs} & \tabincell{c}{Baseline} & \tabincell{c}{TL-OoO} & \tabincell{c}{NUMA} & \tabincell{c}{Cluster} \\
    \hline
    \hline
    Processor~\tnote{1}	& 2$\times$\$1166/3	& 2$\times$\$1166/3	& 4$\times$\$3616/3	& 4$\times$\$1166/3	\\
    \hline
    Memory~\tnote{1}	& 8$\times$\$175/3	& 16$\times$\$175/3	& 16$\times$\$175/3	& 16$\times$\$175/3	\\
    \hline
    \tabincell{c}{Motherboard\\ and Disk}~\tnote{1}
		& \$1000/3	& \$1000/3	& 1.5$\times$\$1000/3~\tnote{2}	& 2$\times$\$1000/3	\\
    \hline
    MEC~\tnote{1}		& -		& 8*100/3		& -		& -		\\
    \hline
    Server power~\tnote{3}
	  	& \$252		& 1.3$\times$\$252	& 1.8$\times$\$252	& 2$\times$\$252	\\
    \hline
    Other costs & \$1325	& \$1325	& 1.5$\times$\$1325~\tnote{4}	& 2$\times$\$1325	\\
    \hline
    \hline
    Total Costs & \$3154	& \$3963	& \$8696	& \$6308	\\
    \hline
    \tabincell{c}{Potential\\Speedup}
		& 1		& $x$		& 2$\times x$		& 2$\times x$	\\
    \hline
    \tabincell{c}{Correction\\Factor}
		& $c=1$		& $c=0.74$		& \tabincell{c}{$c_1=0.76$\\$c_2$ varies}		& $c$ varies	\\
    \hline
  \end{tabular}
{
  \begin{tablenotes}
    \scriptsize
    \item[1] We assume a lifetime of three years.
    \item[2] More processors require a larger motherboard. 
    \item[3] Processors and memory contribute 50\% and 30\%, respectively to server power.
    \item[4] Servers with more processors take more space, increasing data center costs.
  \end{tablenotes}
}
  \end{threeparttable}
  \label{table:cost}
}
}
\end{table}

We compare three ways to extend memory capacity: 
TL coordinates software and MECs, 
NUMA adds more processors, 
and Cluster adds more servers.
The PCIe scheme experiences such slowdowns that we exclude it here.

\paragraph{Cost Model.}
The baseline system has two processors, which is the most cost-effective
configuration, but it only supports one dual-rank RDIMM per channel (a currently
common situation at higher frequencies and a likely continuing trend). 
We choose Intel's mid-end Xeon E5-2650v2 processor with four memory channels and
16GB RDIMM for our comparisons. Our baseline system has 128GB memory in total.

We take costs of server components from the Intel and Amazon websites.
We derive other costs from Barroso and H{\"o}lzle~\cite{barroso2013datacenter}
--- the server cost of a three-year amortization and cost of server power take 50\% and 8\% of
the TCO (total cost of ownership) for a datacenter with mid-end servers\footnote{
Server amortization costs 29.5\% (65.9\%) of TCO for a datacenter
with low-end (high-end) servers, and server power costs represent 14.3\% (3.8\%).}.
Other costs include capital outlay and operating expenses.
Note that for NUMA systems, we must use more expensive processors
that support four cores (here Xeon E5-4650v2). 

We expect that the MEC costs about the same as the LRDIMM buffer,
since both contain two DDRx interfaces. The die area of such a chip is mainly
determined by pin count, rather than logic. We only add
a table and a cache with tens of entries, and thus we do not consume much logic.
We conservatively assume such a MEC costs \$100.

\paragraph{Performance Model.}
We assume that by doubling the memory capacity, performance can at most be improved 
by a factor of $x$. This factor can be quite large in certain cases: 
Graefe et al.~\cite{graefe2014memory} find $x\approx$100 when the extended memory 
capacity can cover an in-memory database's datasets.
NUMA and Cluster also double the number of processors, so
ideal speedup would be 2$\times$.
However, each method brings certain penalties. Our results in
Section~\ref{section:results} shows that TL and NUMA achieve
74\% and 76\% of Ideal performance due to twin-load software and
long latency memory accesses. In addition, NUMA and Cluster also
face the challenge of efficient parallelization.
For applications that are difficult to distribute,
e.g., graph applications, the penalty for the Cluster method can be large. 

\begin{figure}[!t]
\centering
\includegraphics[height=3.5cm]{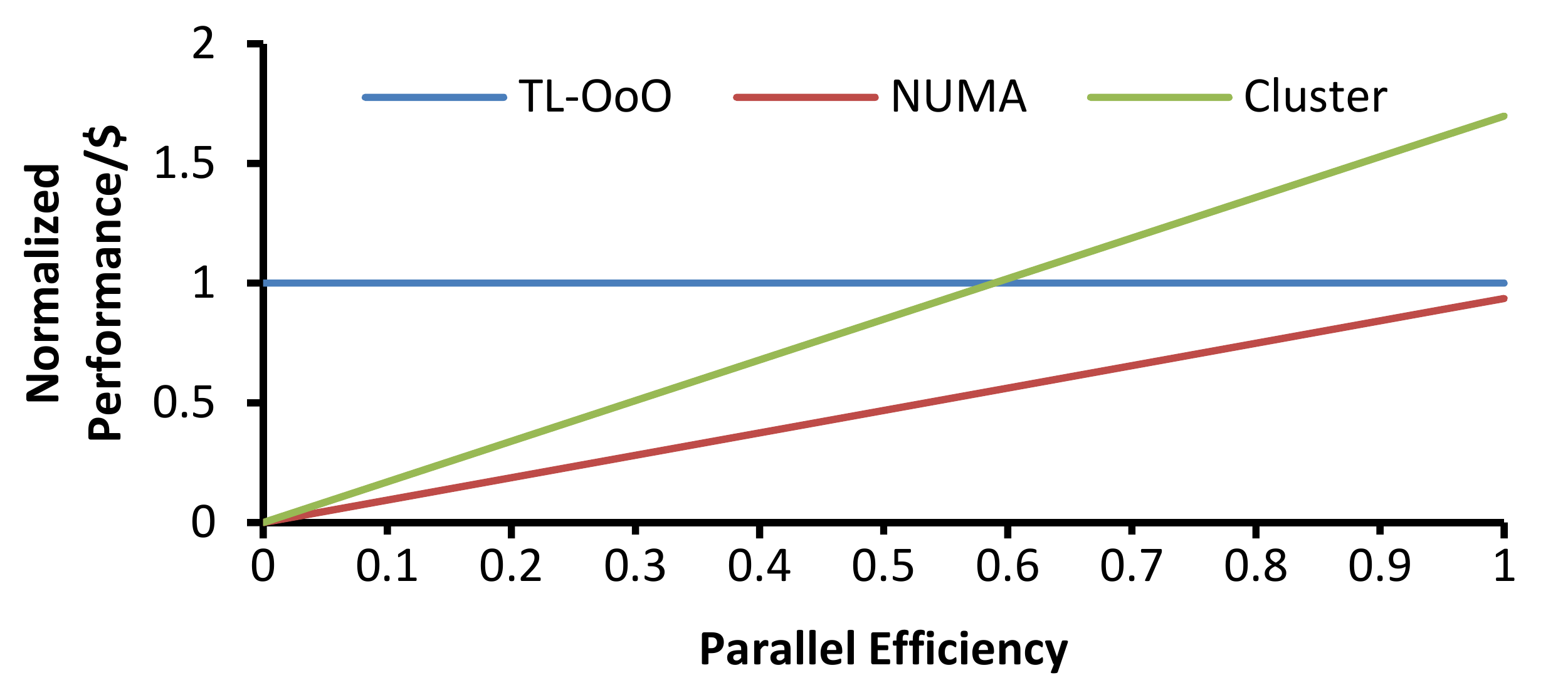}
\caption{Normalized performance per dollar relative to TL-OoO.}
\label{figure:cost}
\end{figure}

\paragraph{Performance per Dollar.}
Table~\ref{table:cost} shows the potential speedups/slowdowns and costs of 
doubling memory capacity for the three systems.
The table shows that relative performance per dollar among 
the mechanisms has no relation to $x$ but rather to the correction
factor $c$ due to twin-load software, cross-processor access latency,
or efficiency of parallel implementation.
Figure~\ref{figure:cost} draws the performance per dollar relative
to the parallel efficiency.
TL can improve performance per dollar by at least 7\%
compared to NUMA when doubling memory capacity. 
At the meantime, TL has the better scalability: the standard
Intel solution only supports up to eight processors, which limits
the system to 4$\times$ the memory capacity.
Clustering has better scalability with respect to memory capacity, but 
it is difficult to scale performance. TL outperforms Cluster whenever 
the distributed application achieves below 60\% of
Ideal performance, which is a challenge for many applications.

\subsection{Comparison with Increased tRL}

To support the larger latency of extended memory, why not just increase the maximum 
latency constraint of JEDEC standard? Since $tRL$, which determines data 
transmission time from memory chip to memory controller can be increased, 
the JEDEC DDRx standard could also adapt to extended memory with larger
latencies. Although this scheme needs a tiny modification to memory controller 
hardware, it is still acceptable. However, according to the DRAM protocol, a memory 
bank will also be held for a longer time, preventing other accesses to that bank. 
This reduces memory bus concurrency, which reduces the benefit of this potential
approach.

We compare TL extended memory to one using a 
single load with increased read latency. We use trace-driven DRAMSim2~\cite{rosenfeld2011dramsim2} 
with dependences between memory instructions~\cite{sanchez2013zsim} 
to simulate the  systems, and we choose more benchmarks than
just those in Table~\ref{table:workloads}.
In the TL system, $tRL$ remains unchanged, but we insert a second load
after each read to cause a row miss. To support latencies greater than 35$ns$,
the second load is delayed for certain time, but does not block following loads
with no data dependence\footnote{
We assume such a mechanism can be realized by manipulating the instruction orders in software.}. 
We compare extra latencies to tolerate of 0-135$ns$. 

\begin{figure}[!t]
\centering
\includegraphics[height=3.5cm]{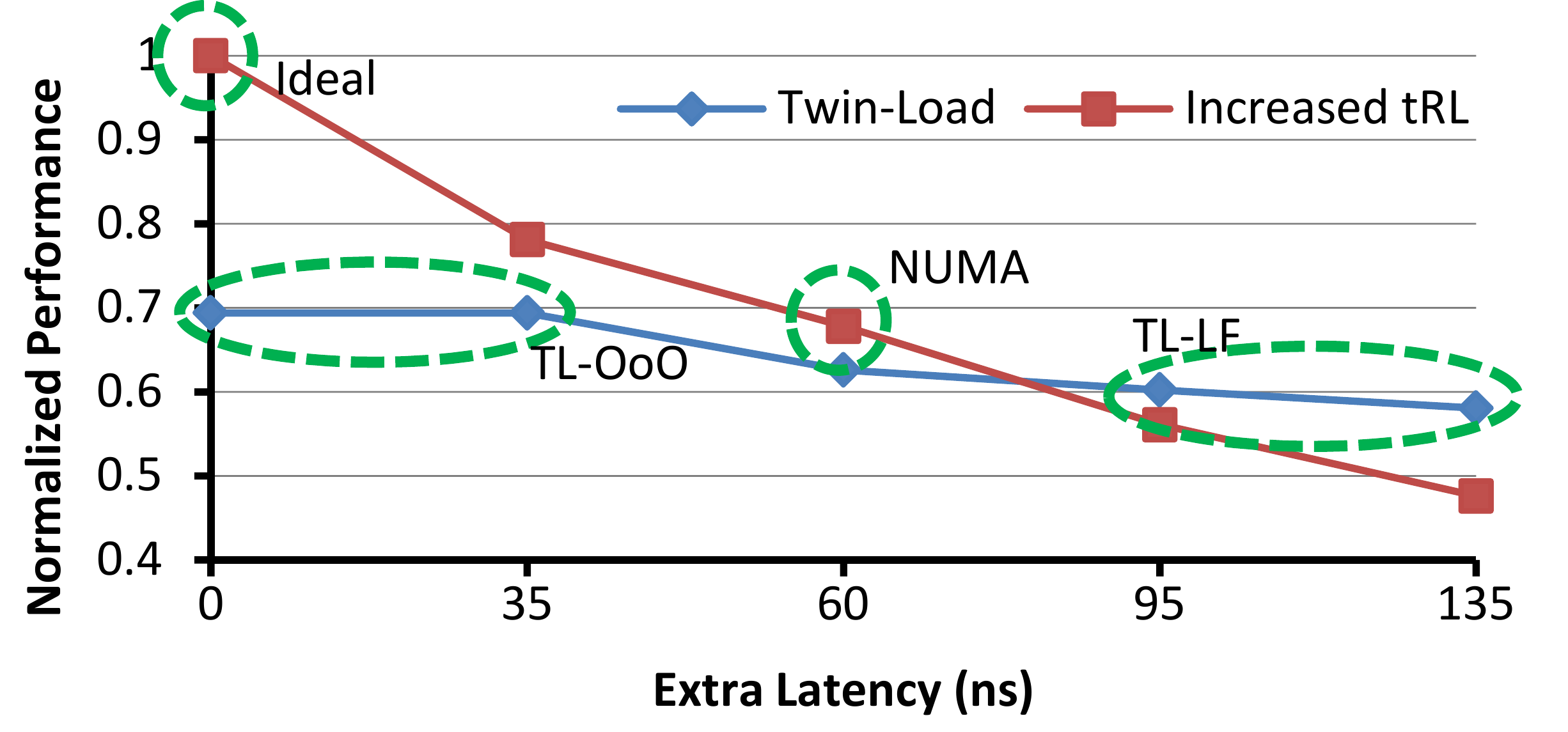}
\caption{Simulated Results of TL vs. Increased $tRL$ (normalized
to tRL=15$ns$ without TL)}
\label{figure:sim}
\end{figure}
  
Figure~\ref{figure:sim} summarizes results. 
The four mechanisms from Section~\ref{section:results} correspond to special
points in the figure and have coincident results expected for TL-LF.
TL-LF
tolerates latencies greater than 100$ns$, but results 
on our emulated prototype are worse than in simulation, mainly because we simulate a
TL mechanism that  does not fence the following loads.
In general, increasing $tRL$ performs better for relatively 
small latencies, but as $tRL$ grows, 
performance degrades faster than for TL because high $tRL$ values limit memory concurrency.
In contrast, the interval between twinned loads can be used to 
execute other memory requests.

\subsection{Comparison with LRDIMM}

Load-Reduced DIMMs are already used to maxmize the server memory capacity. For example the newest
Intel two-socket Xeon E5 server can support up to 1.5TB memory with LRDIMMs, if not considering the cost.

In fact,every buffer-based approach (including MEC) can be considered as reducing the electrical load. 
However, each buffer is typically capable of 2$\sim$4X load reduction, related to frequency. The real 
limitation of LRDIMM is one-layer extension, restricted by CPU’s synchronous DRAM interface and 
access protocol – to be specific, the propagation of more layers violate CPU’s timing constraint.
To the best of our knowledge, commodity processors don’t support cascading of LRDIMM buffers. 
Our proposal breaks such limitation towards more layers and much larger memory capacity using 
software supports, while LRDIMMs still can be used as local memory or extended memory after MEC.

 Since LRDIMM has to put all memory chips within single level , the highest LRDIMM model already uses DDP(Dual-Die-Package) 
 or QDP(Quad-Die-Package), or even 3DS devices. It is not supprised that a single LRDIMM is more expensive than a server CPU.
 While for multi-level MEC extension, more cost-effective RDIMM modules can be incorporated to build a large memory system.
 
 \subsection{Energy}
 
The software overheads of twin-load would increase the energy
consumption compared to the ideal system. For example, the retired
instructions of TL-OoO are 1.64X that of ideal system, indicating
more energy consumption. However, when compared to
a real commodity system, the potential performance improvement
due to twin-load enabled in-memory processing (up to
100X \cite{graefe2014memory}), can actually greatly reduce the total energy consumption
(Energy = Power $\times$ Delay).

\section{Conclusions}

We propose twin-load, a mechanism  to build a lightweight, asynchronous 
data-access protocol that requires no hardware changes on the processor side. 
To achieve this, we coordinate software and the Memory Extending Chips 
on a standard DDRx interface. 
Data access is accomplished by two special loads, the first of which prefetches data
into the top MEC buffer and the second of which brings it into the processor.
Using this mechanism, we can easily attach a multi-layer memory system to 
commodity processors to instantly address the capacity wall problem.
We create an emulation-based software prototype to demonstrate the feasibility
of our proposal. Our best mechanism can achieve 74\% of the performance of
an ideal system with all local memory. 
Our mechanism performs similarly to  NUMA extended memory, but 
delivers much better scalability and performance per dollar.
Twinn-load also outperforms PCIe-based 
systems by several orders of magnitude.

In addition to facilitating easy, cost-effective memory 
extionsions, our mechanism opens opportunities 
to build innovative memory systems on commodity platform using the low-latency,
high-concurrency standard DDRx interface; examples include remote memory pools, 
heterogeneous DRAM/NVM systems, direct remote memory accesses, and even MECs
with integrated accelerators. Relying on open standards and avoiding
changes to processor interfaces enables more system designers -- including
those in academia -- to build production-quality systems.

\bstctlcite{bstctl:etal, bstctl:nodash, bstctl:simpurl}
\bibliographystyle{IEEEtranS}
\bibliography{references}

\end{document}